\documentclass[preprint,aps]{revtex4}
\begin{document}
\title{Excess Demand Financial Market Model.}
\author{Fredrick Michael, John Evans, M.D. Johnson}
\email{mjohnson@ucf.edu, evans@physics.ucf.edu}
\affiliation{Department of Physics, University of Central Florida, Orlando, FL
32816-2385}
\date{July 15, 2002}
\begin{abstract}
Recently we reported on an application of the Tsallis non-extensive
statistics to the S\&P500 stock index \cite{mike1}. There we argued that the
statistics are applicable to a broad range of markets and exchanges where
anamolous (super) diffusion and 'heavy' tails of the distribution are
present, as they are in the S\&P500. We have
characterized the statistics of the underlying security as non-extensive,
and now we seek to generalize to the non-extensive statistics the excess demand 
models of investors that drive the price formation in a market.
\end{abstract}
\maketitle

%%%%%%%%%%%%%%%%%%%%%%%%%%%
In the past decade, there have been many models (\cite{cont1,chowdhury1}%
) proposed that attempt to capture the dynamics and statistics of market
participants that drive the price changes in a market. These range from
minority game models, multi-agent models, and spin models of the bias of
investors. They all have in common the fact that they are models for the
excess demand. The instantaneous excess demand $\phi (t)$ can be defined as
the mismatch in the number of buyers $N_{+}(t)$ and sellers $N_{-}(t)$ for a
number of shares at time $t$. The hallmark for the success of an interacting
investors market model has been the ability of a given model in reproducing
the stylized facts of real markets. These are the heavy tails (power-law) of
the distributions, anomalous (super) diffusion, and therefore statistical
dependence (long-range correlations) of subsequent price changes (\cite
{econophysics}).

Recently the Tsallis non-extensive statistics were applied to analyzing the
price changes and intra-day statistical dynamics of the S\&P500 stock index (\cite
{mike1}). This study characterized the statistics of the price changes as
being well-modeled by the non-extensive statistics. It was also argued that
the statistics are applicable to a broad range of markets and exchanges
where anamolous (super) diffusion and power-law tails of the distribution
are present, as found in the S\&P500 (\cite{econophysics}). \ 

In this paper we examine the demand-side in light of these recent findings
and outline a method by which one can obtain many-investor models within the
context of the Tsallis non-extensive statistics using the maximum entropy
approach. We review the maximum entropy
approach to be utilized. The non-extensive, least-biased probability density
(PDF) $P(z,t)$ of an underlying (say, continuous) observable $z(t)$ is
obtained by maximizing an incomplete information-theoretic measure
equivalent to the Tsallis entropy $S_{q}$ ( \cite{tsallis1,wang1}) and
subject to the known (or assumed) observables of the system as constraints 
\begin{eqnarray}
\left\langle S\right\rangle _{q} &=&S_{q}=-{\frac{1}{1-q}}\left( 1-\int
P(z,t)^{q}\,dx\right) ,  \nonumber \\
\left\langle z\right\rangle _{q} &=&\int z\text{ }P(z,t)^{q}\,dx.
\end{eqnarray}
The maximization of the entropy with $\alpha $ a Lagrange multiplier is 
\begin{equation}
\delta \left\langle S_{q}\right\rangle -\delta \lbrack \alpha \left\langle
z\right\rangle _{q}]\equiv 0,
\end{equation}
and the maximization yields the least biased probability distribution given
the constraints. In the nonextensive statistics this is a $q$-parametrized
power-law Tsallis distribution (\cite{tsallis1}) 
\begin{equation}
P(z,t)=\frac{\{1+\alpha (q-1)\text{ }z\}^{\frac{-1}{q-1}}}{Z_{q}},
\end{equation}
and $q$ is the degree of non-extensivity or equivalently the incompleteness
of the information measure. In the limit of $q\rightarrow 1$ we recover the
Gibbs-Boltzmann extensive statistics and the distribution becomes an
exponential. The normalization is $Z_{q}(t)$, the partition function, and $%
\alpha (t)$ is a possibly time-dependent Lagrange multiplier associated with
the constraints. Again, the constraints will be the known (or assumed)
observables of interest, and are presumed to capture the statistical
behavior of our many-investor system.

We initially wish to model the fluctuating variables of the number of
investors $N_{\pm }(t)$ buying ($+$) and selling ($-$) in the market. We
write the excess demand as $\phi (t)=N_{+}(t)-N_{-}(t)$ and note that it is
proportional to the instantaneous price $x(t).$ That is, we can write
approximately 
\begin{equation}
x(t)=\frac{\phi (t)}{\lambda }=\frac{N_{+}(t)-N_{-}(t)}{\lambda },\text{ }
\label{supply1}
\end{equation}
where $\lambda $ is the market depth, or the excess demand needed to move
the price by one dollar. The market depth is assumed to evolve slowly during
the time scales considered, and will be taken to be a constant. The
observables of interest are the means and variances of the fluctuating
variables. We can write the entropy and observables as statistical averages
about the means

\begin{eqnarray}
\left\langle S\right\rangle _{q} &=&S_{q}=-{\frac{1}{1-q}}\left(
1-\sum\limits_{N_{+},N_{-}=0}^{N}P(N_{+},N_{-},t)^{q}\,\right) ,  \nonumber
\\
\left\langle (N_{+}-\overline{N}_{+})\right\rangle _{q}
&=&0=\sum\limits_{N_{+},N_{-}=0}^{N}(N_{+}-\overline{N}_{+})\text{ }%
P(N_{+},N_{-},t)^{q}\,,  \nonumber \\
\left\langle (N_{+}-\overline{N}_{+})^{2}\right\rangle _{q}
&=&\sum\limits_{N_{+},N_{-}=0}^{N}(N_{+}-\overline{N}_{+})^{2}\text{ }%
P(N_{+},N_{-},t)^{q}\,=\eta _{+}^{2}(t)_{q},  \nonumber \\
\left\langle (N_{-}-\overline{N}_{-})\right\rangle _{q}
&=&0=\sum\limits_{N_{+},N_{-}=0}^{N}(N_{-}-\overline{N}_{-})\text{ }%
P(N_{+},N_{-},t)^{q}\,,  \nonumber \\
\left\langle (N_{-}-\overline{N}_{-})^{2}\right\rangle _{q}
&=&\sum\limits_{N_{+},N_{-}=0}^{N}(N_{-}-\overline{N}_{-})^{2}\text{ }%
P(N_{+},N_{-},t)^{q}\,=\eta _{-}^{2}(t)_{q},
\end{eqnarray}
subject to the additional constraint of normalization $1=\sum%
\limits_{N_{+},N_{-}=0}^{N}P(N_{+},N_{-},t)\,$. We then seek the least
biased probability distribution given these observables. We maximize the
entropy given the constraints 
\begin{equation}
\delta \left\langle S\right\rangle _{q}-\delta \left\{ \beta
(t)[\left\langle (N_{+}-\overline{N}_{+})^{2}\right\rangle _{q}+\left\langle
(N_{-}-\overline{N}_{-})^{2}\right\rangle _{q}]\right\} \equiv 0,
\end{equation}
and here $\beta (t)$ is a time dependent Lagrangian multiplier and can be
shown to be proportional to the inverse of the variance \cite
{mike1}. The maximization yields the power-law least biased
probability distribution 
\begin{equation}
P(N_{+},N_{-},t)=\frac{\left( 1+(q-1)\left[ \beta (t)[(N_{+}-\overline{N}%
_{+})^{2}+(N_{-}-\overline{N}_{-})^{2}\right] \right) ^{\frac{-1}{q-1}}}{%
Z_{q}},
\end{equation}
where $Z_{q}=\sum\limits_{N_{+},N_{-}=0}^{N}P(N_{+},N_{-},t)\,$ is the
partition function and is related to the normalization. We can therefore
obtain all the observables of interest by performing statistical averages
with respect to the distribution. The evaluation of the Lagrange
multiplier(s) proceeds as in the extensive statistics case ( \cite{curado1}) 
\begin{eqnarray}
-\frac{\partial }{\partial \beta }\frac{Z_{q}^{1-q}-1}{1-q} &=&\left\langle
(N_{+}-\overline{N}_{+})^{2}\right\rangle _{q}+\left\langle (N_{-}-\overline{%
N}_{-})^{2}\right\rangle _{q}  \nonumber \\
&=&\eta _{+}^{2}(t)_{q}+\eta _{-}^{2}(t)_{q},  \label{observe1}
\end{eqnarray}
and in the limit as $q\rightarrow 1$ we recover the standard extensive
statistics expression $-\frac{\partial lnZ_{1}}{\partial \beta }%
=\left\langle \vartheta \right\rangle _{1}$.

The demand-side model then provides us with a statistical description of the
fluctuating variables ($N_{+},N_{-}$), the difference of which can be
related to the price by $x(t)=\frac{N_{+}(t)-N_{-}(t)}{\lambda }$. This
model has the power-law behavior of real markets as a consequence of the
pseudo-additive nature of the entropy of the subsystems. That is to say, the
number of investors buying and selling at time $t$ in the 'subsystems' of
buyers and sellers are not statistically independent, and the incremental
changes in time of the number of buyers and sellers will be correlated in
time. It is also known that the price changes $dx(t)$ exhibit anomalous
diffusion and correlations in time and therefore by $dx(t)=\frac{d\phi (t)}{%
\lambda }$ so will the change in the excess demand $d\phi (t)$. The
composition of the entropy of the two subsystems must be written from the
joint probability decomposition (suppressing the time parameter) 
\begin{equation}
P(N_{+},N_{-})=P(N_{+}\mid N_{-})P(N_{-})
\end{equation}
which gives the pseudo-additive entropy 
\begin{eqnarray}
S_{q}(N_{+},N_{-}) &=&S_{q}(N_{-})+S_{q}(N_{+}\mid
N_{-})+(1-q)S_{q}(N_{-})S_{q}(N_{+}\mid N_{-}),  \nonumber \\
S_{q} &=&-ln_{q}P=-\frac{P^{1-q}-1}{1-q}.
\end{eqnarray}
It can be shown (\cite{rajagopal1}) that the Tsallis entropy satisfies this
condition, and the resulting probability will be of the power-law form
derived above.

In order to simplify the solution, let us symmetrize the range of the
distribution. We define $n_{\pm }=N_{\pm }-\frac{N}{2}$ such that the
variable $n_{\pm }$ now has the symmetrical range $-\frac{N}{2}\leq n_{\pm
}\leq \frac{N}{2}$ and the excess demand becomes $\phi =n_{+}-n_{-}$. We
next pass to the continuum limit of the number of investors. Alternatively,
we could have started our derivation from this approximation. Also, due to
the observation that most trades involve a small fraction of the overall
number of investors ( $N_{\pm }-\overline{N}_{\pm }\ll N$) we relax the
range to $N\rightarrow \infty $. The distribution $P(n_{+},n_{-},t)$ is now
taken to be symmetric and continuous, and is given by a similar power-law
form as before 
\begin{equation}
P(n_{+},n_{-},t)=\frac{\left( 1+(q-1)\left[ \beta (t)[(n_{+}-\overline{n}%
_{+})^{2}+(n_{-}-\overline{n}_{-})^{2}\right] \right) ^{\frac{-1}{q-1}}}{%
Z_{q}}.
\end{equation}
We integrate to obtain the partition function 
\begin{eqnarray}
Z_{q}(t) &=&\int\limits_{-\infty }^{\infty }\int\limits_{-\infty }^{\infty
}P(n_{+},n_{-},t)dn_{+}dn_{-} \\
&=&\frac{\pi \text{ }}{\beta (t)(2-q)},
\end{eqnarray}
where the range of $q$ must now be restricted to $1<q<2$ to insure
normalization. We note that $\beta (t)Z_{q}(t)=c_{q\text{ }}$, a time
independent constant of the process dependent on the non-extensivity
parameter $q$. This is useful in that we can relate the Lagrange multiplier
parameter $\beta (t)$ (and therefore the inverse variance) to the partition
function for all times considered. The range of $q$ is chosen as $1\leq
q\leq q_{max}$ consistent with the requirements that the distribution is
normalizable for all times and that the regular variance remain finite.

%\[
%\FRAME{itbpF}{247.625pt}{148.25pt}{0pt}{}{}{Figure }{\special{language
%"Scientific Word";type "GRAPHIC";display "USEDEF";valid_file "T";width
%247.625pt;height 148.25pt;depth 0pt;original-width 61pt;original-height
%61pt;cropleft "0";croptop "1";cropright "1";cropbottom "0";tempfilename
%'GZAALR06.wmf';tempfile-properties "XPR";}}
%\]

The regular variances $\eta _{\pm }^{2}(t)$ can be computed readily from the
distribution 
\begin{eqnarray}
\left\langle (n_{\pm }-\overline{n}_{\pm })^{2}\right\rangle
&=&\int\limits_{-\infty }^{\infty }\int\limits_{-\infty }^{\infty }(n_{\pm }-%
\overline{n}_{\pm })^{2}P(n_{+},n_{-},t)dn_{+}dn_{-} \\
&=&\frac{1}{2\text{ }\beta (t)\text{ }(3-2q)},  \nonumber
\end{eqnarray}
and the range of $q$ must be further restricted to $1<q<1.5$ to assure
convergence of the integral and therefore a finite variance.

The power-law distribution $P(n_{+},n_{-},t)$ has a time evolution that
satisfies a two dimensional Fokker-Planck partial differential equation ( 
\cite{zanette1,tsallis1}), and we assume a linear drift term $\alpha (n_{\pm
})=a_{\pm }$ $n_{\pm }$ consistent with the linear drift of the price
distribution (\cite{mike1}) 
\begin{eqnarray}
\frac{\partial }{\partial t}P(n_{+},n_{-},t) &=&-\frac{\partial }{\partial
n_{+}}[\alpha (n_{+})\text{ }P(n_{+},n_{-},t)]-\frac{\partial }{\partial
n_{-}}[\alpha (n_{-})P(n_{+},n_{-},t)]  \nonumber \\
&&+\frac{D}{2}\left( \frac{\partial }{\partial n_{+}^{2}}+\frac{\partial }{%
\partial n_{-}^{2}}\right) [P^{2-q}(n_{+},n_{-},t)].
\end{eqnarray}
This subsequently implies the underlying stochastic differential equations ( 
\cite{gardiner1}) for the number of buyers and sellers

\begin{eqnarray}
dn_{+} &=&\alpha (n_{+})dt+\sqrt{DP^{1-q}(n_{+},n_{-},t)}\text{ }dW_{+}(t), 
\nonumber \\
dn_{-} &=&\alpha (n_{-})dt+\sqrt{DP^{1-q}(n_{+},n_{-},t)}\text{ }dW_{-}(t).
\label{sde1}
\end{eqnarray}
Here $dW_{\pm }(t)=\xi _{\pm }(t)dt$ , where $W_{\pm }(t)$ is the Wiener
process and $\xi _{\pm }(t)$ is a delta correlated noise such that $%
\left\langle \xi _{\sigma }(t)\xi _{\sigma }(t^{\prime })\right\rangle
=\delta _{\sigma \sigma ^{\prime }}\delta (t-t^{\prime })$. If we examine
the change in the excess demand we obtain the relationship to the change of
price
\begin{eqnarray*}
d\phi  &=&dn_{+}-dn_{-} \\
&=&\lambda \text{ }dx(t)
\end{eqnarray*}
The solutions of these SDE's can be obtained by using Ito's change of
variables and performing the integration over time
\begin{equation}
n_{\pm }(t)=n_{\pm }(t_{o})e^{a_{\pm
}(t-t_{o})}+\int\limits_{t_{o}}^{t}e^{a_{\pm }(t-t^{\prime })}\sqrt{%
DP^{1-q}(n_{+},n_{-},t)}dt^{\prime },
\end{equation}
and we thus have the desired result for the formation of the instantaneous
excess demand $\phi (t)$, and therefore the instantaneous price $x(t)$%
\begin{eqnarray}
\phi (t) &=&n_{+}(t)-n_{-}(t)  \nonumber \\
&=&\lambda \text{ }x(t).
\end{eqnarray}
We note that the solutions of these stochastic equations depends on the
knowledge of the time evolution of the probability $P(n_{+},n_{-},t)$. One
must then solve the Fokker-Planck equation simultaneously with the SDE's.
Equivalently one can obtain the time evolution of $\beta (t)$ analytically
and therefore the evolution of $P(n_{+},n_{-},t)$ as outlined in (\cite{mike1}).   

%We numerically simulate the stochastic evolution(s). The random trajectory
%of the excess demand $\phi (t)$ stochastic process is plotted in Fig.1. with 
%$q=1.49$, $\sqrt{D}=0.15$ and $a_{\pm }=0.1$ for $800$ minutes and with $%
%\Delta t=1$min.  The time evolution of $\beta (t)$, related to the inverse
%variance is plotted in Fig.3. and displays a similar stretched exponential 
%decay as the inverse of the variance obtained from the analysis of high 
%frequency price data from the S\&P500 (\cite{mike1}).    
%
%\[
%\FRAME{itbpF}{272.4375pt}{162.5625pt}{0in}{}{}{Figure }{\special{language
%"Scientific Word";type "GRAPHIC";display "USEDEF";valid_file "T";width
%272.4375pt;height 162.5625pt;depth 0in;original-width
%64.75pt;original-height 64.75pt;cropleft "0";croptop "1";cropright
%"1";cropbottom "0";tempfilename 'GZAALR05.wmf';tempfile-properties "XPR";}}
%\]
%
%\[
%\FRAME{itbpF}{255.125pt}{166.3125pt}{0in}{}{}{Figure }{\special{language
%"Scientific Word";type "GRAPHIC";display "USEDEF";valid_file "T";width
%255.125pt;height 166.3125pt;depth 0in;original-width 64.75pt;original-height
%64.75pt;cropleft "0";croptop "1";cropright "1";cropbottom "0";tempfilename
%'GZAALR07.wmf';tempfile-properties "XPR";}}
%\]
%
\bigskip  
The stochastic differential equations (SDEs) in Eq.(\ref{sde1}) can be seen
to be a non-extensive SDE generalizations of the Cont and Bouchaud model (
\cite{cont1}), in the continuum approximation. Moreover, they are of the
statistical feedback form (\cite{mike1,borland1}), in that the microscopic
stochastic dynamics are coupled to the macroscopic probability distribution $%
P(n_{+},n_{-},t)$. The microscopic stochastic equations and the macroscopic
Fokker-Planck equation are known to be equivalent descriptions of a given
random process. In this case of anomalous diffusion of the excess demand,
the statistical dependence of subsequent demand changes are explicitly
represented in the SDE. Hence this statistical model of the excess demand
reproduces the stylized facts of real markets, namely power-law
distributions and anomalous diffusion as a consequence of the
pseudo-additivity of the system entropy.

We have in this article generalized a model of excess demand and price
formation to the case of the non-extensive statistics of C. Tsallis. We
state the need to go beyond the extensive Gibbs-Boltzmann statistics in
modeling real markets from the point of view of the pseudo-additivity of the
entropy of statistically dependent subsystems, and the decomposition of the
non-factorizable joint probability. This approach to statistically dependent
subsystems and statistically correlated variables has been applied directly
to the price statistics of real markets (\cite
{mike1,mike2,borland1,borland2,ramos1}) and shows promise in modeling the
stylized facts of these markets, such as the power-law behavior and
anomalous diffusion of price changes. We are working to examine this
model given data sets of investor demand and market depths from a real
market (such as the S\&P500) that is known to exhibit the anomalous
diffusion and power-law behavior of price changes. We are also recasting 
this model of investor demand in terms of spins (bias), and moving beyond 
this number statistics model to an interacting many-investor spin model of 
markets.  

M.D. Johnson and F. Michael acknowledge support from the NSF through grant
number DMR99-72683.

\end{document}